\documentclass[manuscript]{aastex}

\newcommand{\chan}{\textit{Chandra}}

\shorttitle{X-ray Study of CTB~87}
\shortauthors{Matheson, Safi-Harb \& Kothes}

\begin{document}

\title{
X-ray Observations of the Supernova Remnant CTB~87 (G74.9+1.2): An Evolved Pulsar Wind Nebula}
\author{H. Matheson\altaffilmark{1} and S. Safi-Harb\altaffilmark{2}}
\affil{Department~of~Physics~\&~Astronomy, University~of~Manitoba, Winnipeg, Manitoba, R3T~2N2, Canada}

\and

\author{R. Kothes}
\affil{Dominion Radio Astrophysical Observatory, National~Research~Council~Herzberg, \\P.O. Box 248, Penticton, British Columbia, V2A~6J9, Canada}

\altaffiltext{1}{matheson@physics.umanitoba.ca}
\altaffiltext{2}{Canada Research Chair, samar@physics.umanitoba.ca}
\altaffiltext{3}{roland.kothes@nrc-cnrc.gc.ca}

\begin{abstract}
Pulsar wind nebulae (PWNe) studies with the \textit{Chandra X-ray Observatory} have opened a new window to 
address the physics of pulsar winds,
zoom on their interaction with their hosting supernova remnant (SNR) and interstellar medium, and identify their powering engines.
We here present 
a new 70~ks, plus an archived 18~ks, \chan\ ACIS observation of the SNR~CTB~87 (G74.9+1.2), classified as a PWN with unusual radio properties and poorly studied in X-rays.
We find that the peak of the X-ray emission is  
clearly offset from the peak of the radio emission by $\sim$100\arcsec\ and located at the southeastern edge of the radio nebula. 
 We detect a point source - the putative pulsar - at the peak of the X-ray emission and study its spectrum separately from the PWN.  This new point source, CXOU~J201609.2+371110,
 is surrounded by a compact nebula displaying a torus-like structure and possibly a jet.
A more extended diffuse nebula is offset from the radio nebula,
extending from the point source to the northwest for
$\sim$250\arcsec.  
The spectra of the point source, compact nebula and extended diffuse nebula are all well described by a power law model with a photon index of 1.1 (0.7$-$1.6), 1.2 (0.9$-$1.4)
and 1.7 (1.5$-$1.8), respectively, for a column density $N_{\rm{H}}$~=~1.38 (1.21$-$1.57)~$\times$~10$^{22}$~cm$^{-2}$ (90\% confidence).
The total X-ray luminosity of the source is
$\sim$1.6~$\times$~10$^{34}$~erg~s$^{-1}$ at an assumed distance of 6.1~kpc, with $\sim$ 2\% and 6\% contribution from the point source and compact nebula, respectively.
The observed properties suggest that CTB~87 is an evolved ($\sim$5$-$28~kyr) PWN, with the extended radio emission likely a `relic' PWN, as in Vela-X and G327.1$-$1.1. 
To date, however, there is no evidence for thermal X-ray emission from this SNR, and the SNR shell is still missing, suggesting expansion into a 
low-density medium ($n_0 < 0.2~D^{-1/2}_{6.1}$~cm$^{-3}$),  
likely caused by a stellar wind bubble blown by the progenitor star.
\end{abstract}


\section{Introduction}
Pulsar wind nebulae (PWNe) are the bubbles inflated by the relativistic winds of neutron stars as they interact with their surroundings.
Their radio to X-ray emission
is believed to result from synchrotron radiation of the high-energy particles injected by 
a rotation-powered neutron star in the presence of a strong magnetic field.
As such, PWNe represent an ideal laboratory to study the physics of neutron stars and particle acceleration.  They are also good indicators for the presence of an active rotation-powered pulsar;
see~\citet{Gaensler2006} for a detailed
review on the structure and evolution of PWNe.
While the Crab Nebula has been viewed to represent the prototype PWN, a growing list of PWNe show properties unlike the Crab: namely, different spectral properties (spectral index, spectral break in the radio) and different morphology.  These are referred to in the literature as ``plerions of the second kind'' or non-Crab-like plerions \citep[see][]{Woltjer1997}, a class that we have been targeting with combined radio and X-ray studies; 
see~\citet{SafiHarb2012} for a recent review.
This paper addresses, using X-ray observations, the nature of CTB~87; the least studied object (at least in X-rays) among this growing class of PWNe. 
In a subsequent paper (Kothes et al. in preparation), we present a dedicated study of the remnant at radio wavelengths.

CTB~87 (G74.9+1.2)\footnote{Main properties summarized at www.mrao.cam.ac.uk/surveys/snrs/\\snrs.G74.9+1.2.html~\citep{Green2009}}$^,$\footnote{High-energy properties summarized at www.physics.umanitoba.ca/snr/SNRcat/\\SNRrecord.php?id=G074.9p01.2 \citep{Ferrand2012}} has been classified as a plerionic type SNR based on its centrally-filled morphology, lack of evidence of a shell, and linearly polarized radio flux~\citep{Dickel1975, Duin1975}.  In the radio, CTB~87 has a size of 8\arcmin$\times$6\arcmin\ and a flux density of 9~Jy at 1~GHz~\citep{Green2009}. 
Early X-ray observations with the \textit{Einstein} satellite also identified CTB~87 as a plerionic SNR, with an X-ray luminosity nearly 100 times weaker than the Crab Nebula in the $0.15-3$~keV band~\citep{Wilson1980}.  
Newer sensitive radio studies indicated that the central bright structure has a relatively steep radio spectrum, with a spectral index $\alpha$ = 0.5  
($S_{\nu}~\sim~\nu^{-\alpha}$), which is unusual for a plerionic SNR.
In addition, larger diffuse emission, $\sim$16\arcmin\ in size, has been identified and characterized by a much flatter spectrum which is more consistent of a plerionic SNR (Kothes et al. 2003; Kothes et al. in preparation).
\citet{Sitnik2010} suggested that CTB~87 is located in the Cygnus Arm, 
and detected a very faint optical shell with 20\arcmin\ diameter in the vicinity of CTB~87.  This optical shell has not been however confirmed and the author concludes that the small amount of data prevents them from drawing a firm conclusion.
Locating CTB~87 in the Cygnus Arm puts the remnant at a much smaller distance than that estimated using HI data from the Canadian Galactic Plane Survey ($6.1 \pm 0.9$~kpc,~\citet{Kothes2003}), a distance that
implies a location within the Perseus spiral arm.
An \textit{EGRET} source has been suggested to be possibly associated with CTB~87, although~\citet{Halpern2001} argued that the SNR is too weak and far away (using an old distance estimate of 12~kpc) to be the most likely candidate.  
The revised kinematic distance to CTB~87 of 6.1~kpc
weakened this argument, further increasing interest in CTB~87 as a potential $\gamma$-ray source.
Other $\gamma$-ray studies of the area near CTB~87 with \textit{MILAGRO}  led to the detection of an unresolved source,
MGRO~J2019+37~\citep{Abdo2007}, 
which was further studied and resolved with \textit{VERITAS}.
The \textit{VERITAS} source, VER~J2016+372, showed an absence of variability throughout the observations and a spectrum similar to other PWNe previously detected in the VHE band~\citep{Aliu2011}.
Finally, \textit{FERMI} detected a high-energy source, 2FGL~2015.6+3709, which was found to be likely variable and therefore a physical association with CTB~87 was deemed improbable~\citep{Nolan2012}.  

Here we present a detailed study of the X-ray observations of
CTB~87. 
 Preliminary results have been published in~\citet{SafiHarb2011} and~\citet{SafiHarb2012}.  
This paper is focused on the observations acquired with the \textit{Chandra X-ray Observatory}, with a total exposure time of $\sim$87~ks.
We also briefly present an unpublished archival observation with the \textit{Advanced Satellite for Cosmology and Astrophysics (ASCA)}, primarily to compare the overall spectrum of the nebula with that obtained with \chan.  Section~\ref{section:obs} discusses the observations and data processing. In Section~\ref{section:images}, we present the observed X-ray morphology and show the first evidence for a compact nebula, with a torus/jet-like structure, as well as a diffuse nebula that is offset from the radio nebula. These structures surround a point source, the putative neutron star powering CTB~87. 
 Section~\ref{section:spectra} presents the spectral analysis, including the first spatially resolved spectroscopic analysis. We show that the spectra from the point source and the compact and diffuse nebulae are well described by a power law with a hard photon index, and that the index generally steepens away from the point source.  In Section~\ref{section:disc} we discuss the PWN morphology, properties, and stage of evolution, and infer the properties of the putative neutron star powering the PWN.  We propose that CTB~87 is most likely an evolved PWN, with overall multi-wavelength properties similar to the PWNe in the SNRs Vela and G327.1$-$1.1.

\section{Observations}
\label{section:obs}

\object{CTB~87} (G74.9+1.2) was observed with \chan\ on two occasions.  An 18~ks observation was performed with ACIS-S3 (2001 July 8, obsID 1037)  
and a 70~ks observation (2010 January 16, obsID 11092) was performed with ACIS-I3.   
The \chan\ data was processed with CIAO v4.3 and CALDB v4.4.1~\citep{Fruscione2006}\footnote{http://cxc.harvard.edu/ciao/}.  The level 1 event files were reprocessed to new level 2 event files, using \textit{acis\_process\_events} (correct for CCD charge transfer inefficiencies\footnote{http://cxc.harvard.edu/ciao/why/cti.html}, remove bad pixels) and \textit{dmcopy} (remove bad grades and status, filter for good time intervals and energy).    
The effective exposures after filtering were 17.8~ks (obs 1037) and 69.3~ks (obs 11092), yielding a total effective exposure of 87.1~ks.

For image creation, the two observations were merged after applying the aspect solution to each observation (\textit{reproject\_events}).  
\textit{wavdetect} was used to detect point sources and \textit{dmfilth} was used to remove the sources and fill the holes with a Poisson distribution calculated from a background region surrounding each source.  \textit{asphist}, \textit{mkinstmap}, and \textit{mkexpmap} were used to create exposure maps for each CCD, which were then summed.  
 \textit{dmimgthresh} was used to remove pixels with very low exposure times and the image was divided by the exposure map  
to produce the final exposure corrected image.  This procedure was repeated for various energy bands and the resulting images are presented in Section~\ref{section:images}.

Spectra were extracted from each observation separately, and simultaneously fit using \textit{XSPEC} v12.6.0\footnote{http://xspec.gsfc.nasa.gov}. 
\textit{specextract} was used to extract spectra and response files from regions of interest.  The regions used and resulting spectra are discussed in Section~\ref{section:spectra}.  When discussing diffuse emission, each larger region excludes events contained in the interior regions, as well as any counts from point sources detected within the diffuse emission.  Spectra were grouped using \textit{dmgroup}, with the minimum number of counts per bin listed in Table 2.  All uncertainties quoted on model parameters are to 90\% confidence level.  For calculations requiring a distance estimate, we assume $D=6.1D_{6.1}$~kpc~\citep{Kothes2003}.

CTB~87 was also observed on 1995 May 29 with \textit{ASCA} (sequence\# 53041000).  It was imaged with the GIS for 53 ks but was out of the field of view of the SIS.  Data was processed with \textit{XSELECT} (HEASoft v5.3.1).  We present the \textit{ASCA} results only for the purpose of comparing the spectral result for the PWN to the global fit with \chan.
The X-ray observations of CTB~87 presented in this paper are summarized in Table~\ref{table:observations}.

\section{Imaging}
\label{section:images}

\subsection{Nebula Imaging}
\label{section:nebulaimage}

\begin{figure}
\epsscale{1.00}
\plottwo{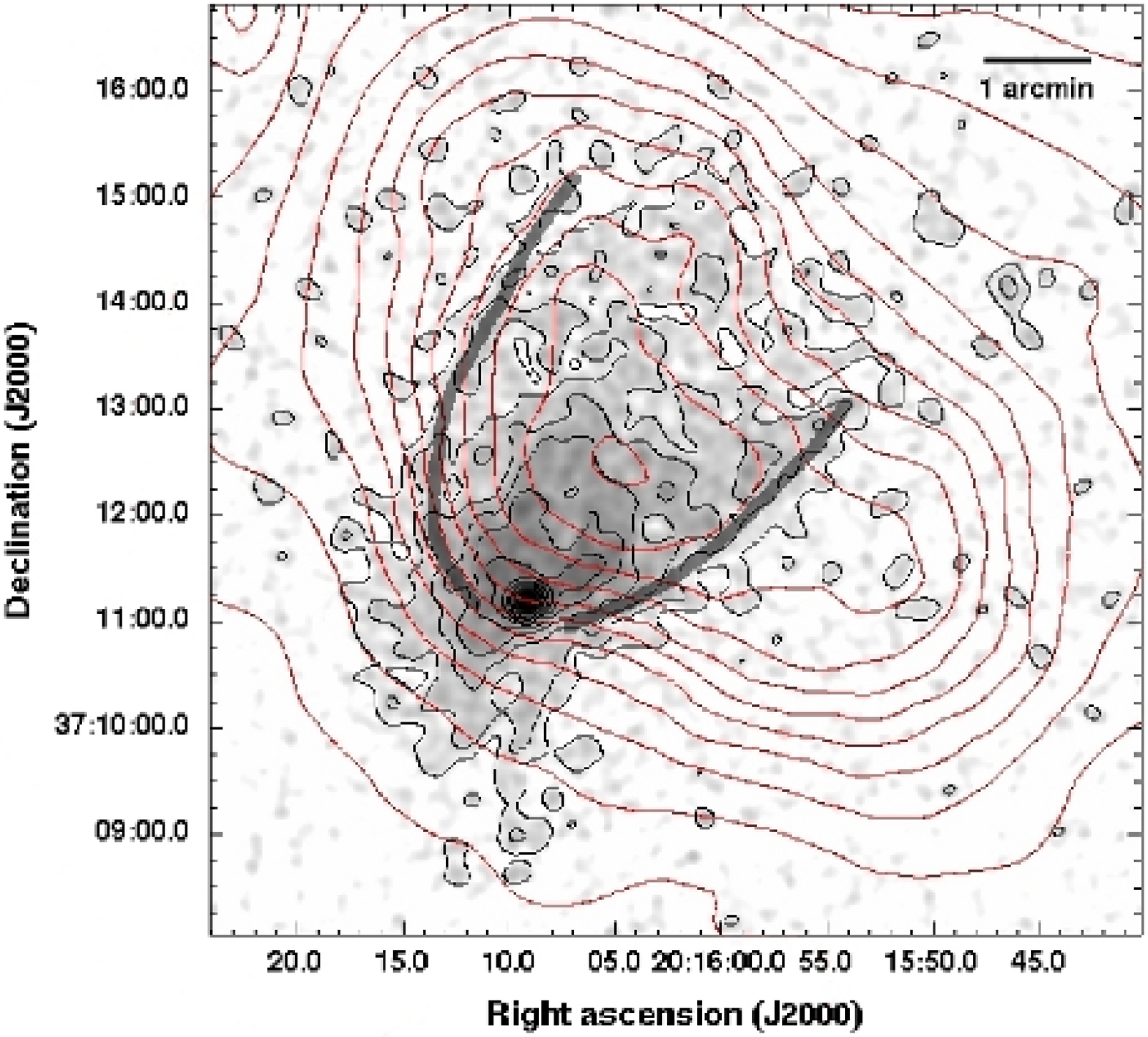}{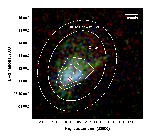}
\caption{\textbf{Left:} 0.3$-$7.0~keV \chan\ 87.1~ks image of CTB~87.  Logarithmic scaling is used and the image is smoothed with a 2D Gaussian ($\sigma=3\arcsec$).  Black contours highlight the X-ray morphology.  Overlaid arcs highlight the brightening along the edge of the X-ray PWN.  Red contours correspond to the radio 21 cm emission (CGPS data) and define the brightness temperature at the following levels: 93.0, 86.0, 78.9, 71.9, 64.9, 57.8, 50.8, 43.8, 36.8, 29.7, 22.7, 15.7, 8.6, 1.6 Kelvin.
 \textbf{Right:}  RGB \textit{Chandra} image (red = 0.3$-$2.0 keV, green = 2.0$-$4.0 keV, blue = 4.0$-$7.0 keV).  Regions overlaid were used to extract spectra of the diffuse emission.  Inner regions were excluded when extracting a spectrum for an outer region.  The fit to all of the diffuse emission (Section~\ref{section:nebulaspectra}, Table~\ref{table:spectra}) includes emission from the Inner + Mid + Outer regions.  The outer annulus shows the background region used for the diffuse emission spectra.}
\label{figure:pwn}
\end{figure}

Figure~\ref{figure:pwn} (left panel) shows the combined \chan\ observations 1037 and 11092, after removing point sources, filling the holes, and exposure correcting the image, as described in Section~\ref{section:obs}.  The energy range covered is 0.3$-$7.0~keV 
(events below 0.3~keV and above 7.0~keV were excluded as the background dominates at these energies) 
and the net exposure time is 87.1~ks. 
The image was smoothed with a 2D Gaussian ($\sigma$~=~3\arcsec) and 21~cm contours are overlaid, using data acquired with the \textit{Dominion Radio Astrophysical Observatory} as part of the Canadian Galactic Plane Survey~\citep{Taylor2003}.
 The entire image is 525\arcsec~$\times$~525\arcsec.  The PWN is $\sim$200\arcsec~$\times$~300\arcsec\ overall with brighter emission to the southeast.  The 
total observed X-ray flux of the nebula is 1.8~$\times$~10$^{-12}$~erg~cm$^{-2}$~s$^{-1}$ ($\sim$0.3~cts~s$^{-1}$).  
The peak of the X-ray emission is seen in the southeast of the nebula, with a point source clearly visible at the peak (Section~\ref{section:ptimage}).  The X-ray peak is located near the southeastern edge of the bright radio nebula, offset from the center of the nebula, with the diffuse emission extending 100\arcsec, 100\arcsec, 250\arcsec, and 60\arcsec\ from the peak to the SE, NE, NW, and SW, respectively.  We observe arcs at the edges of the X-ray nebula (highlighted on Figure~\ref{figure:pwn}), extending from the southeast near the point source, back toward the radio peak. 
The nature of these features is discussed in Section~\ref{section:disc}.

Figure~\ref{figure:pwn} (right panel) is a color image created by dividing the data into three bands,  exposure correcting each, and smoothing ($\sigma$~=~8\arcsec) each before combining.  The red, green and blue colors
correspond to the 0.3$-$2.0~keV, 2.0$-$4.0~keV, and 4.0$-$7.0~keV bands, respectively.
The contrast of each image was adjusted to reduce the background and bring out the PWN emission.  We see the hardest (bluest) emission near the bright point source (the putative pulsar, Section~\ref{section:ptimage}).  The arcs along the edge of the PWN (highlighted on Figure~\ref{figure:pwn}, left panel) appear as possible bow shock structures and are real:
the signal-to-noise ratio of the eastern arc is 12.8 while the signal-to-noise ratio of the western arc is 11.5.
In Section~\ref{section:disc} we discuss the nature of these features.

\subsection{Compact Object Imaging}
\label{section:ptimage}

\begin{figure}
\epsscale{1.0}
\plottwo{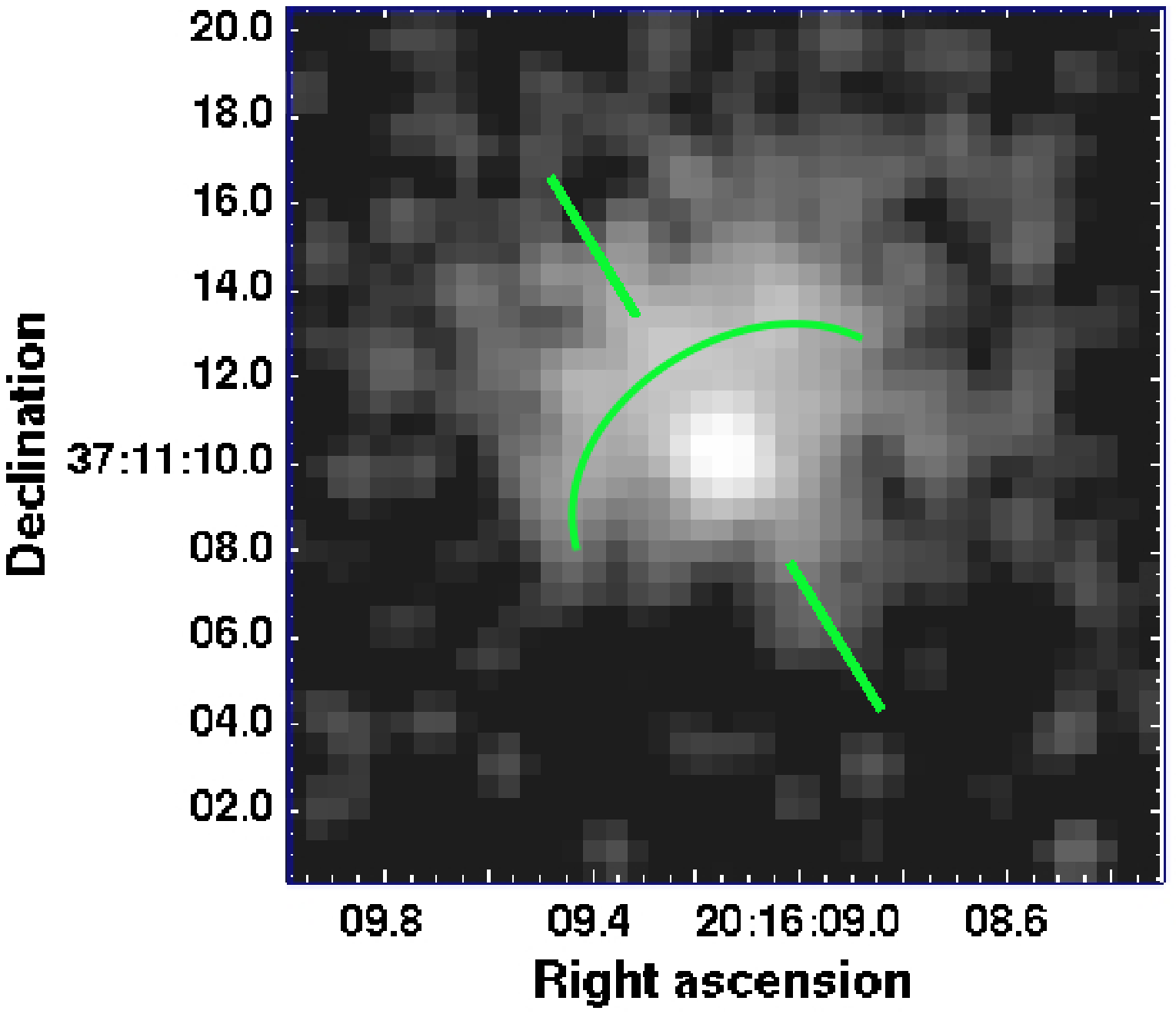}{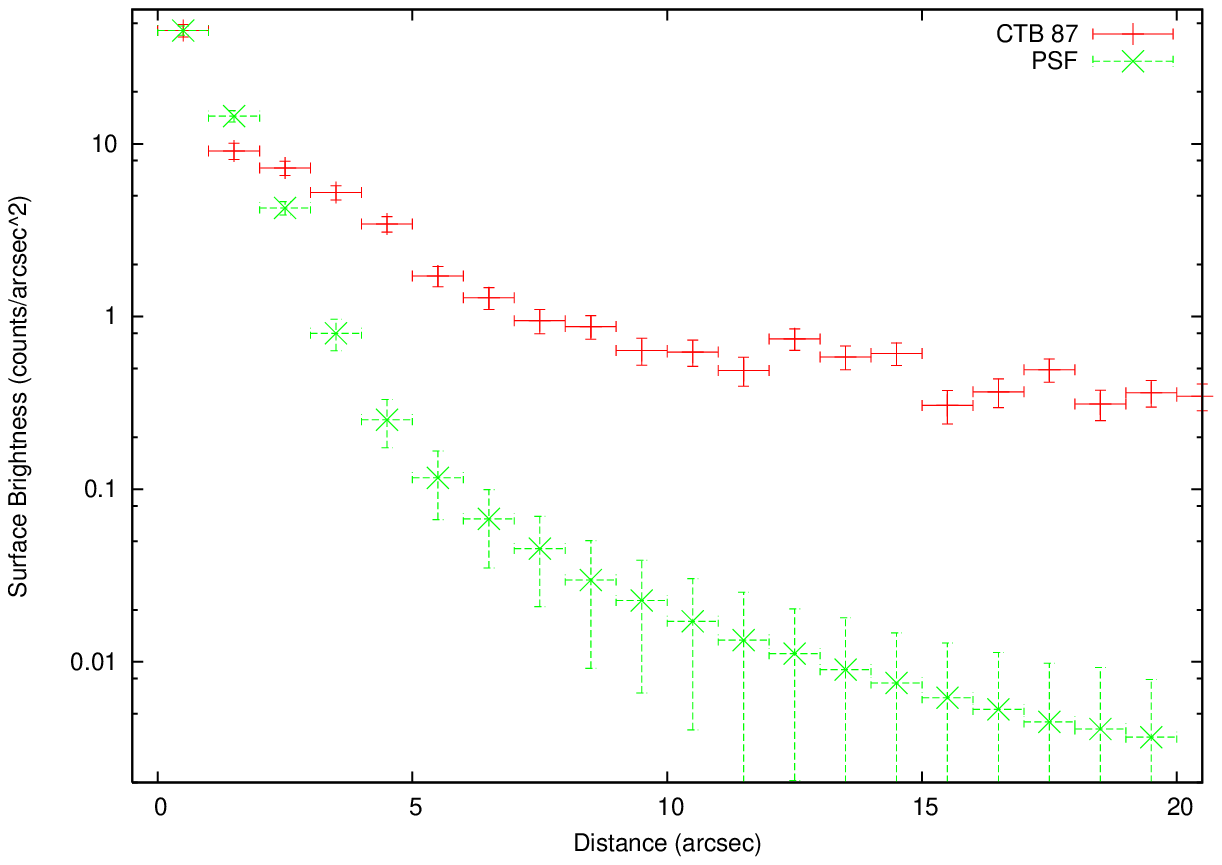}
\caption{\textbf{Left:} The \textit{Chandra}-detected point source located at $\alpha$(2000)=20$^{\rm{h}}$16$^{\rm{m}}$09\fs2, $\delta$(2000)=+37\arcdeg11\arcmin10\farcs5 and the surrounding compact nebula.  Image is 20\arcsec$\times$20\arcsec, 87.1~ks (obs. 1037 and 11092), 0.3$-$7.0~keV, exposure corrected and smoothed using a Gaussian with $\sigma$~=~1\arcsec.  The arc highlights a torus candidate and the lines highlight candidate jets.  
\textbf{Right:} (Red) Radial profile of the \chan\ data centered on the peak of the X-ray emission. (Green) Radial profile of a PSF with the same peak energy as the data and at the same off-axis angle.}
\label{figure:pointsource}
\end{figure}

Thanks to \chan's superb resolution, a point source - the putative pulsar - is found at the southeastern edge of the bright radio nebula  with the following coordinates: $\alpha$(2000)=20$^{\rm{h}}$16$^{\rm{m}}$09\fs2, $\delta$(2000)=+37\arcdeg11\arcmin10\farcs5. 
We refer to this new \chan\ point source as CXOU~J201609.2+371110.\footnote{Despite the lack of pulsations from this source, we refer to it hereafter as `pulsar' or `neutron star'.} 
A point spread function (PSF) was created using \textit{mkpsf} for an energy of 3~keV (characteristic of the source's energy histogram) and an off-axis angle of 2\farcm38 (the location of the point source in observation 11092) and found to have a FWHM of 1\farcs15.  We note that this is the positional uncertainty of the point source.  
Figure~\ref{figure:pointsource} (left) shows a 20\arcsec~$\times$~20\arcsec\ region, centered on the point source and combining both \textit{Chandra} observations (as for Figure~1).  This exposure corrected image was smoothed using a Gaussian with $\sigma$~=~1\arcsec\
and covers the energy range 0.3$-$7.0 keV.  We see evidence of an arc-like (torus) structure to the northeast of the point source with $\sim$10\arcsec\ diameter, and jet-like features extending $\sim$5\arcsec\ to the northeast and southwest.  Such
high-resolution features are expected to form close to the termination shock, due to  the deposition of the neutron star's wind energy into its surroundings~\citep[see e.g.][]{Gaensler2006}.

Figure~\ref{figure:pointsource} (right) shows a radial profile created from the observational data, 
along with a radial profile of the PSF.  The data is consistent with a point source to a radius of $<$2\arcsec, beyond which the data clearly has a higher surface brightness than a point source.  This excess emission we term the `compact nebula' and is comprised of the candidate torus and jets, as seen in the left panel of Figure~\ref{figure:pointsource}.  We also tested the source extent using \textit{srcextent}.  The source observed size was 2\farcs05, the PSF observed size was 1\farcs15, and the estimated intrinsic size was 1\farcs69, confirming that there is an extended source surrounding the point source.

\begin{figure}
\epsscale{1.00}
\includegraphics[width=2.1in]{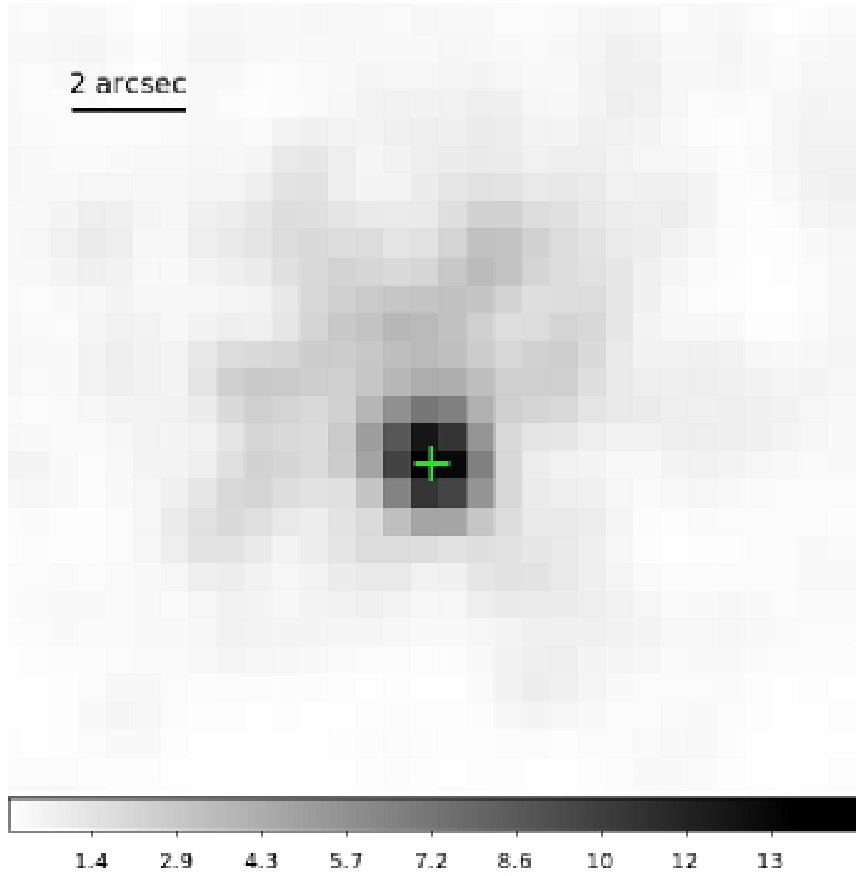}
\includegraphics[width=2.1in]{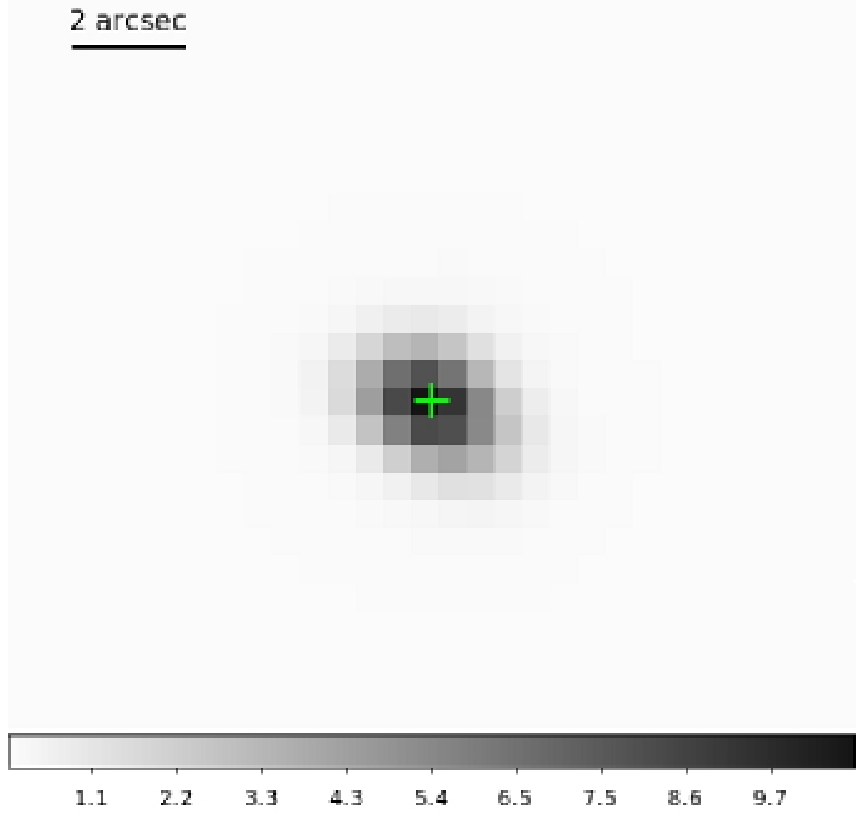}
\includegraphics[width=2.1in]{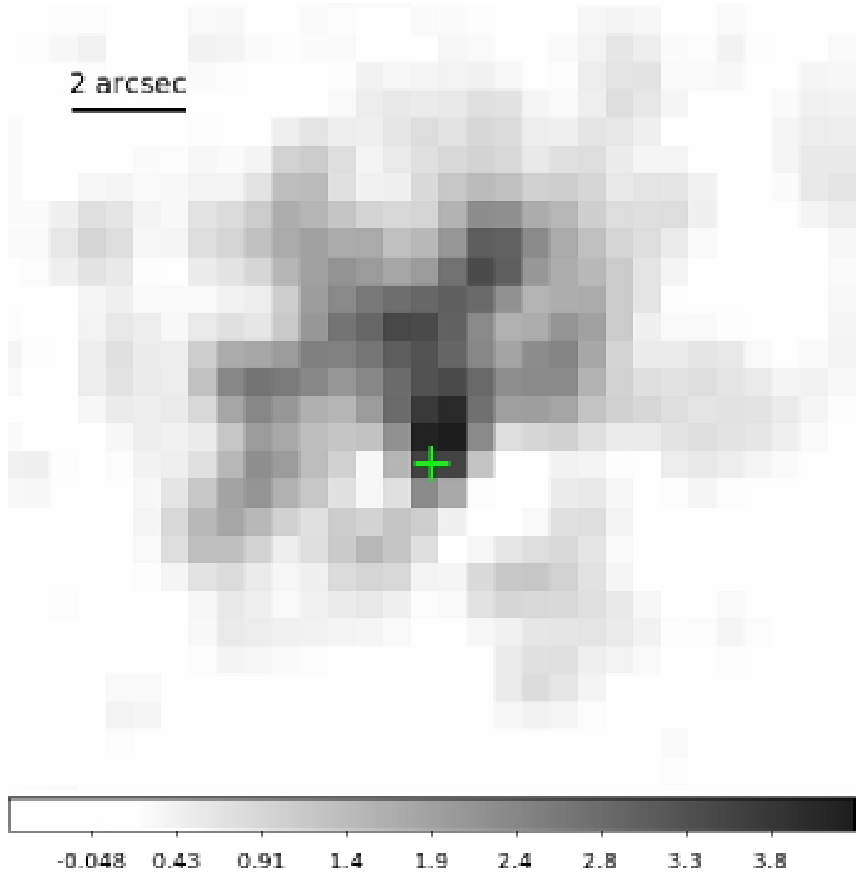} 
\caption{\textbf{Left:}  The \chan\  observation 11092 shown at the peak of the X-ray emission and smoothed with $\sigma$~=~1\arcsec. \textbf{Middle:} 2D model of the point source emission, convolved with the PSF.  \textbf{Right:} Residuals after subtracting the model from the data, smoothed with $\sigma$~=~1\arcsec.  All three panels cover the same area.  The cross marks the peak of the X-ray emission and is at the same location in all images.  See Section~\ref{section:ptimage} for details of PSF creation, model, and the fitted parameters.}
\label{figure:residuals}
\end{figure}

The PSF described above was then used as a convolution kernel in $Sherpa$\footnote{http://cxc.harvard.edu/sherpa4.4/index.html} to model the point source emission and determine the morphology of the excess emission.  Figure~\ref{figure:residuals} shows the data, the model, and the residuals resulting from the fit. The model consisted of a constant background level plus a 2D Gaussian with a FWHM of 1\farcs1 and an amplitude of 117 counts/pixel.  We see a large excess 
in the residual map 
to the northeast of the point source location, seeming again to have the torus-jet morphology as suggested in Figure~\ref{figure:pointsource}.

\section{Spectroscopic Analysis}
\label{section:spectra}

\subsection{Compact Object and Compact Nebula's Spectra}
\label{section:ptspectra}

\begin{figure}
\plottwo{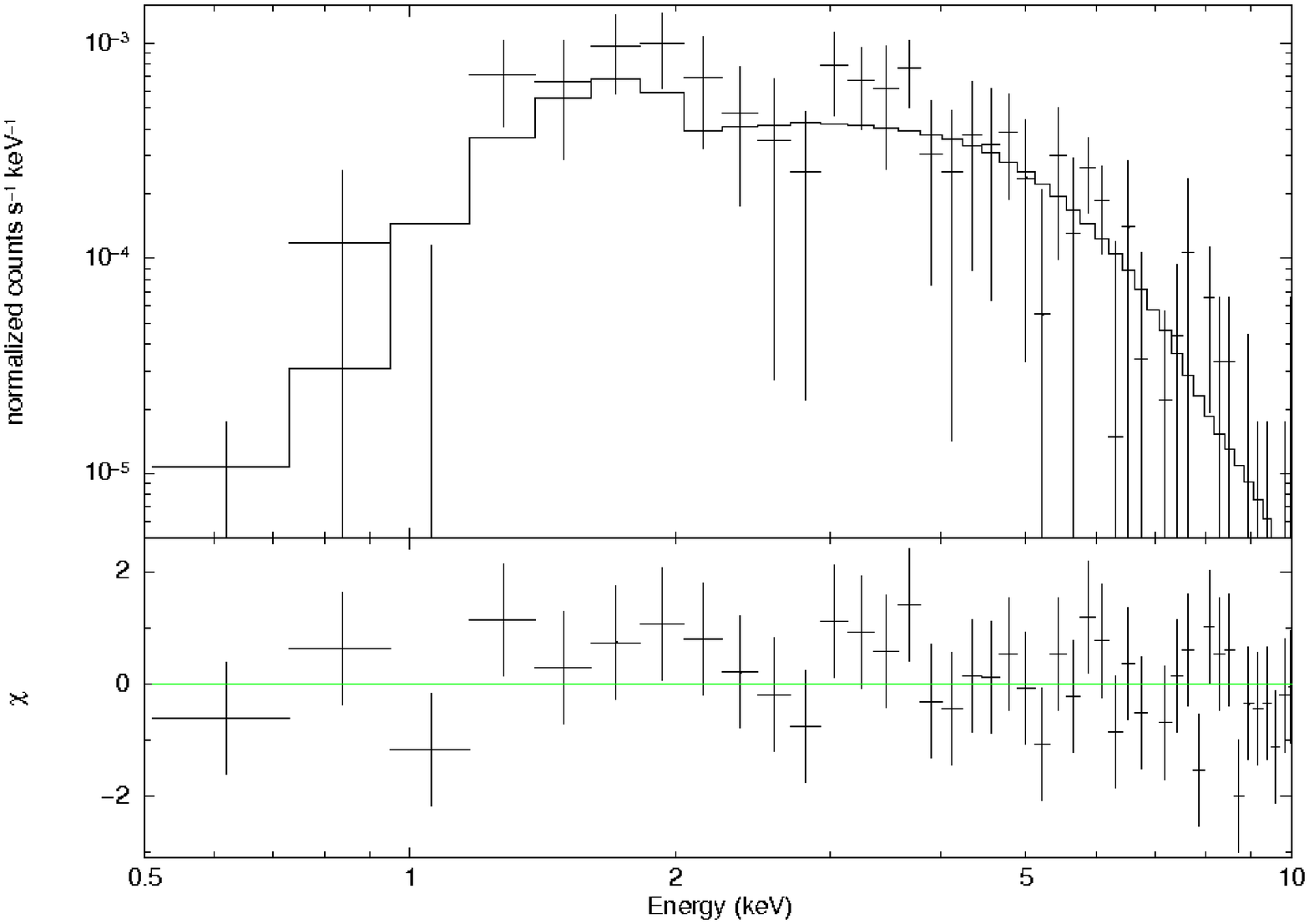}{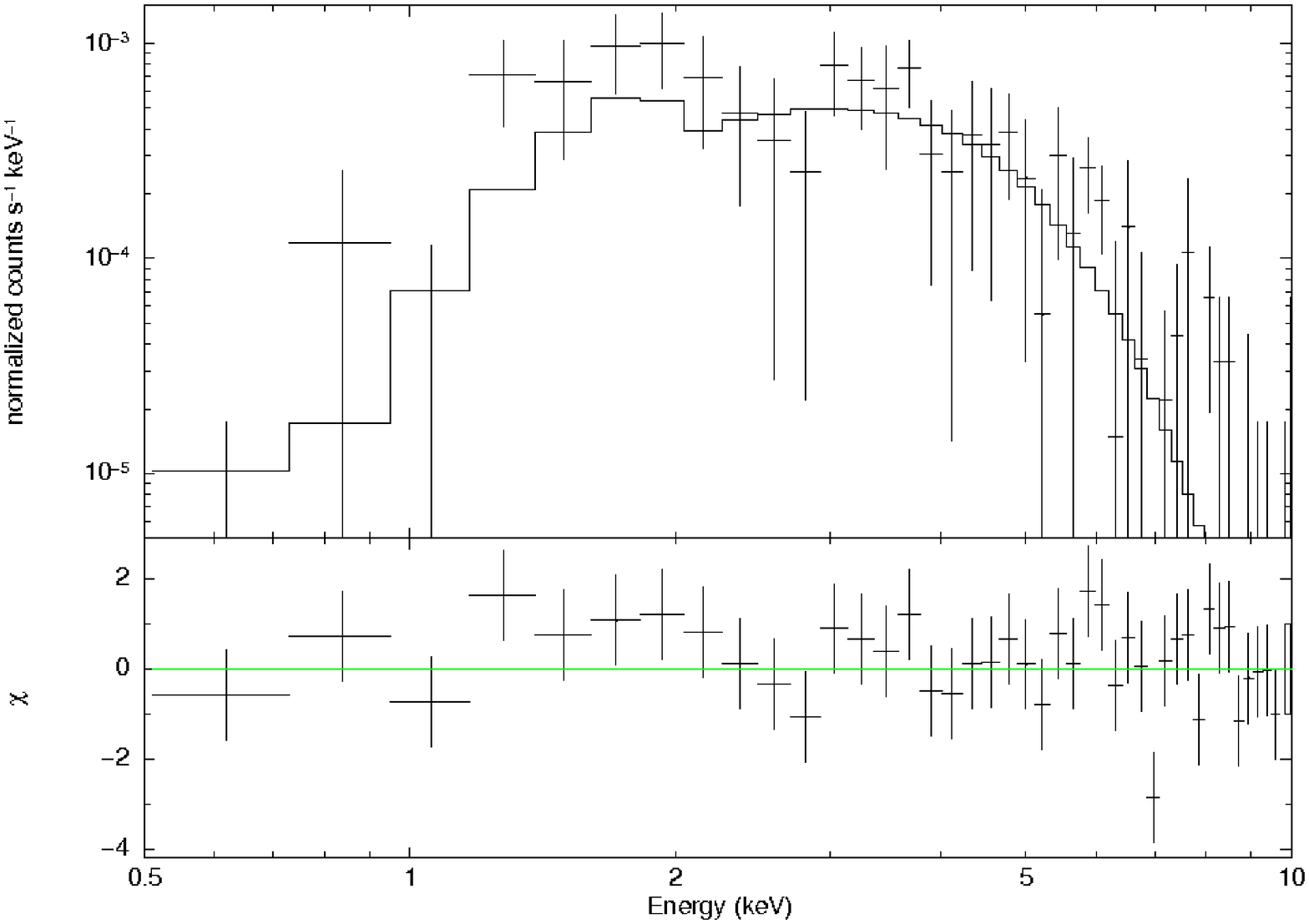}
\caption{Spectrum of the point source (0\arcsec$-$2\arcsec, centered on the X-ray peak).  \textbf{Left:} Absorbed power law fit to the putative pulsar.  \textbf{Right:} Absorbed blackbody fit to the putative pulsar.  Model parameters are listed in Table~\ref{table:spectra} and the fits are discussed in Section~\ref{section:ptspectra}.}
\label{figure:pointspectrum}
\end{figure}

To extract the spectrum of the point source, we used a circular region centered at CXOU J201609.2+371110 and with a radius~=~2\arcsec.
The background spectrum was extracted using an annulus surrounding the source with radius~=~2\arcsec$-$4\arcsec. 
Given the small number of counts (see Table~2), we freeze the column density, $N_{\rm{H}}$, to the best value of 1.4~$\times$~10$^{22}$~cm$^{-2}$ obtained from fitting the total diffuse emission  (see Section~\ref{section:nebulaspectra}).
We attempted fitting with an absorbed power law model and an absorbed blackbody model, as would be expected from a neutron star.
Fitting with the absorbed power law model 
yields a hard photon index of $\Gamma$~=~1.1~(0.7$-$1.6)
 and an absorbed flux of $F_{\rm{abs,~0.3-10.0~keV}}$~=~5.2~$\times$~10$^{-14}$~erg~cm$^{-2}$~s$^{-1}$ ($\chi^2_{\nu}$~=~0.91; see Table~2).
 The corresponding X-ray luminosity is $L_{\rm{0.3-10.0~keV}}$~=~3.0~$\times$~10$^{32}$~$D_{6.1}^2$~erg~s$^{-1}$ or $L_{\rm{2-10~keV}}$~=~2.5~$\times$~10$^{32}$~$D_{6.1}^2$~erg~s$^{-1}$.  The left panel of Figure~\ref{figure:pointspectrum} shows the data and fitted model, with the spectra from the two observations grouped together for plotting purposes only.  We note that allowing the column density to vary gives a column density of $N_{\rm{H}}$~=~1.6~(0.2$-$3.0)~$\times$~10$^{22}$~cm$^{-2}$ ($\chi^2_{\nu}$~=~0.93), consistent with that of the diffuse nebula, but due to the low number of counts the parameters are much less constrained.

Fitting with an absorbed blackbody model we find a similarly suitable fit 
($\chi^2_{\nu}$~=~0.99, 
Figure~\ref{figure:pointspectrum}, right) with a temperature $kT$~=~1.1~(0.8$-$1.5)~keV, an absorbed flux of $F_{\rm{abs,~0.3-10.0~keV}}$~=~3.2~$\times$~10$^{-14}$~erg~cm$^{-2}$~s$^{-1}$, and a luminosity $L_{\rm{0.3-10.0~keV}}$~=~1.8~$\times$~10$^{32}$~$D_{6.1}^2$~erg~s$^{-1}$ or $L_{\rm{2-10~keV}}$~=~1.6~$\times$~10$^{32}$~$D_{6.1}^2$~erg~s$^{-1}$.  However, the temperature is unreasonably high and, when combined with the luminosity, implies a neutron star radius of $\sim$0.03~km which is unreasonably small. 

Attempting a two component model (power law plus blackbody) was not possible given the poor statistics.
Therefore we favor the power law model fit, which is also consistent with the X-ray emission seen from other neutron stars powering PWNe, such as PSR~J2022+3842 in G76.9+1.0~\citep[$\Gamma$~=~1.0~$\pm$~0.2,][]{Arzoumanian2011}, and PSR~J1833-1034 in G21.5-0.9~\citep[$\Gamma$~=~1.14~$\pm$~0.07,][]{Matheson2010}.

\begin{figure}
\plotone{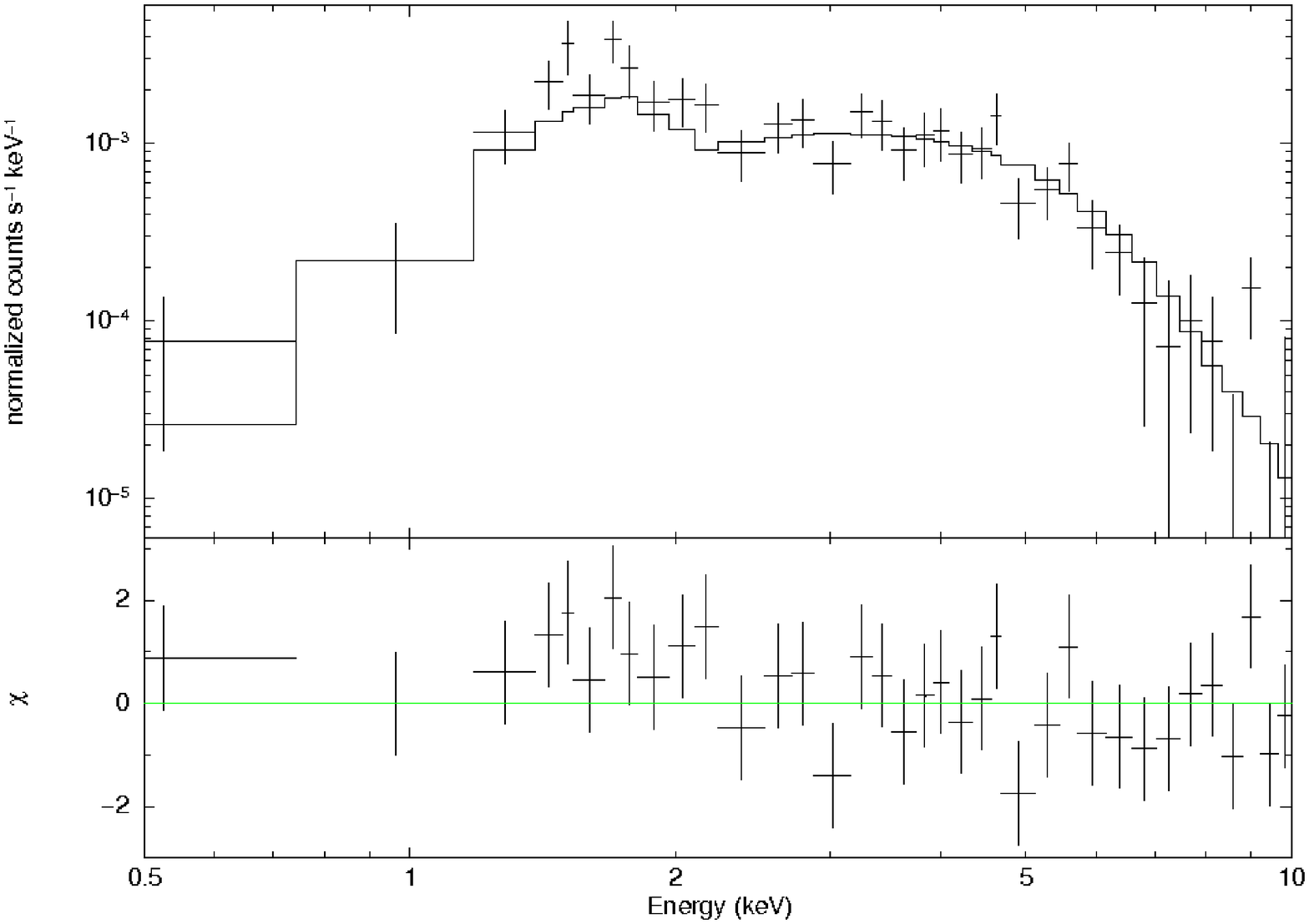}
\caption{Spectrum of the compact nebula surrounding the point source (2\arcsec$-$10\arcsec, centered on the X-ray peak), fit with an absorbed power law model.}
\label{figure:compactnebspectrum}
\end{figure}

We then studied the emission from the compact
nebula by extracting a spectrum from an annulus with radius 2\arcsec$-$10\arcsec, using a larger annulus with radius 10\arcsec$-$20\arcsec\ 
 as the background spectrum.  This compact nebula, like the point source, has a hard spectrum with $\Gamma$~=~1.2~(0.9$-$1.4) and an X-ray luminosity of  $L_{\rm{0.3-10.0~keV}}$~=~8.7~$\times$~10$^{32}$~$D_{6.1}^2$~erg~s$^{-1}$ or $L_{\rm{2-10~keV}}$~=~6.5~$\times$~10$^{32}$~$D_{6.1}^2$~erg~s$^{-1}$.  The best-fit spectral parameters are summarized in Table~\ref{table:spectra} and the compact nebula spectrum is shown in Figure~\ref{figure:compactnebspectrum}.

\subsection{Diffuse Nebula's Spectrum}
\label{section:nebulaspectra}

To study the properties of the extended diffuse emission, we extracted a spectrum of the extended nebula shown in Figure~\ref{figure:pwn} (right image), omitting the inner 10\arcsec\ radius compact nebula.
A power law model provides a good fit, as would be expected from a PWN, 
with the following best-fit parameters (see Table~2):   $N_{\rm{H}}$~=~1.38~(1.21$-$1.57)~$\times$~10$^{22}$~cm$^{-2}$,   $\Gamma$~=~1.68~(1.54$-$1.84),
$\chi^2_{\nu}$~=~1.01, $F_{\rm{abs,~0.3-10.0~keV}}$~=~2.1~$\times$~10$^{-12}$~erg~cm$^{-2}$~s$^{-1}$, 
$L_{\rm{0.3-10.0~keV}}$~=~1.6~$\times$~10$^{34}$~$D_{6.1}^2$~erg~s$^{-1}$, and $L_{\rm{2-10~keV}}$~=~9.6~$\times$~10$^{33}$~$D_{6.1}^2$~erg~s$^{-1}$.

\begin{figure}
\plottwo{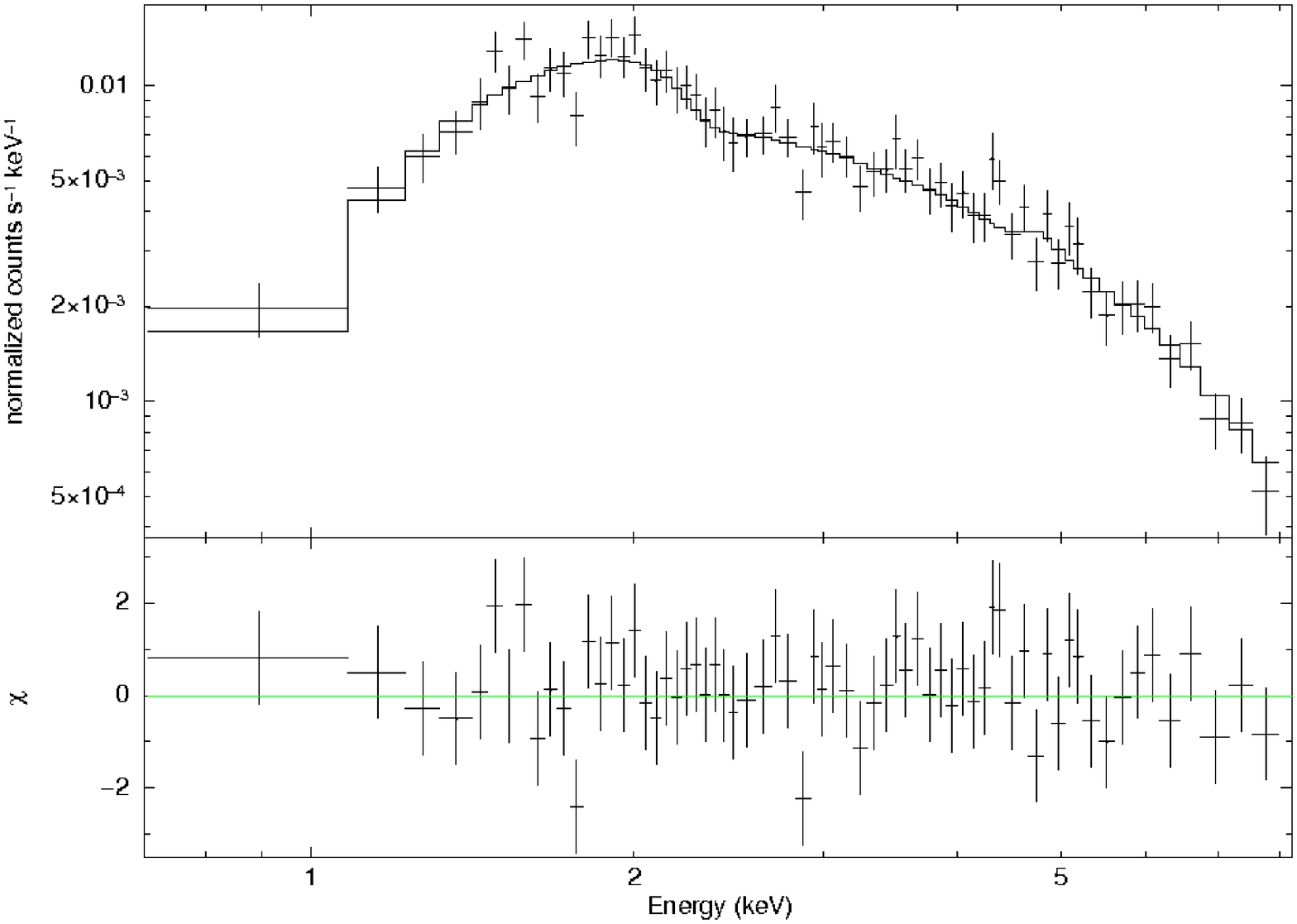}{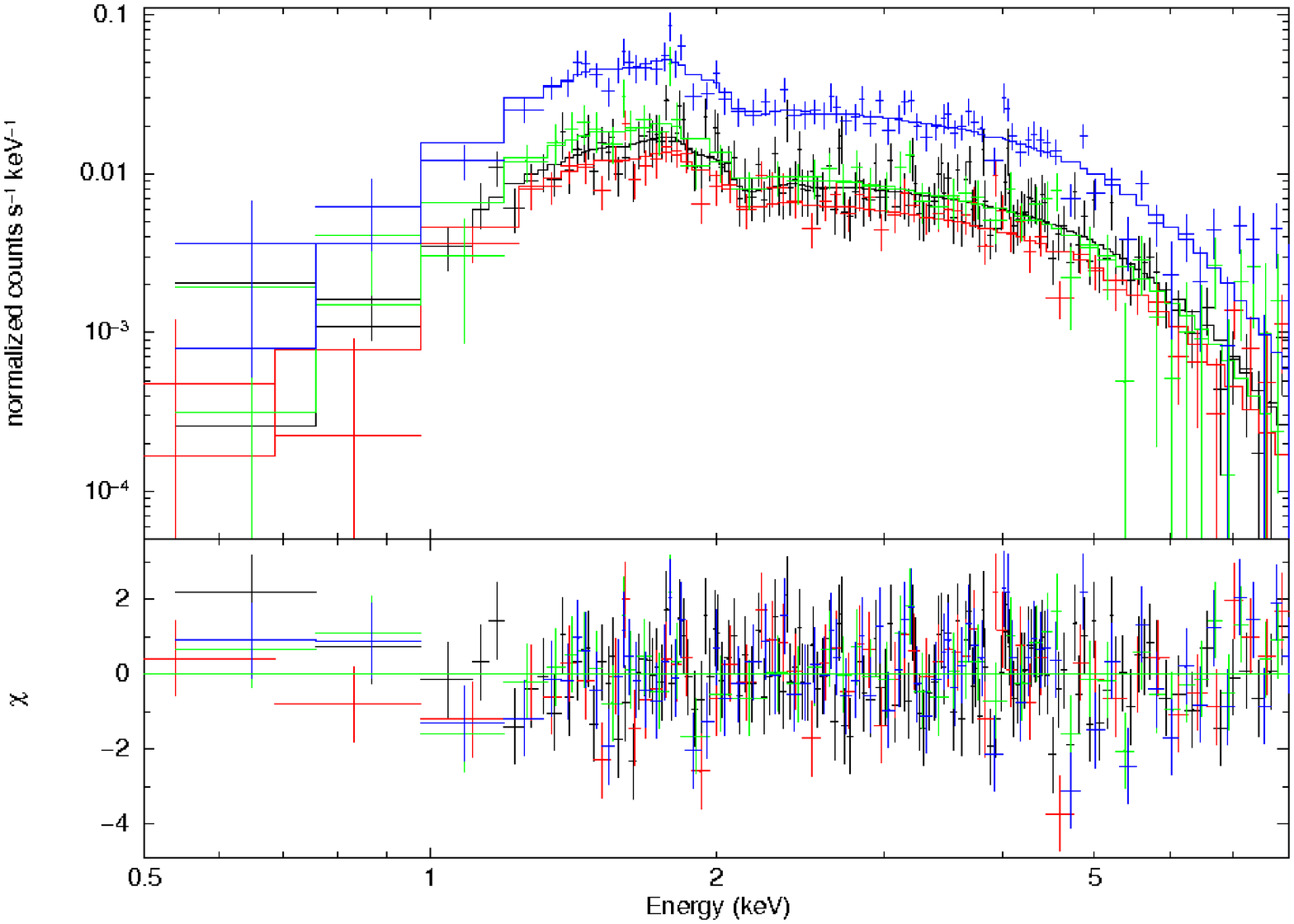}
\caption{\textbf{Left:} \textit{ASCA} spectrum of CTB~87.  Emission includes the point source and all of the diffuse emission, as resolved spectroscopy was not possible.  See Section~\ref{section:nebulaspectra} for the spectral parameters.  \textbf{Right:} \textit{Chandra} spectra of the PWN diffuse emission (regions shown in Figure~\ref{figure:pwn}).  Spectra from obs.~11092 and obs.~1037 were grouped for plotting purposes only.  Black = inner PWN.  Red = mid PWN.  Green = outer PWN.  Blue = entire PWN = inner + mid + outer.  The solid lines are the best fit (wabs*power) to the data, with the parameters listed in Table~\ref{table:spectra}.}
\label{figure:pwnspectrum}
\end{figure}

The above fit is consistent with the results we find from archived \textit{ASCA} GIS data.  The point source in CTB~87 was not resolved in the \textit{ASCA} observation and so we fit the entire nebula, a circular region with a radius of 3\farcm4, with a background spectrum extracted from the same detector, outside the diffuse emission.  Using GIS-2 and GIS-3 data (CTB~87 was not visible in the SIS data), we accumulate 3213 counts after background subtraction (compared to 10136 net counts with \textit{Chandra}).  Fitting GIS-2 and GIS-3 data simultaneously with a power law model in the 0.8$-$8.0 keV energy band, we find that $N_{\rm{H}}$~=~1.14~(0.94$-$1.36)~$\times$~10$^{22}$~cm$^{-2}$, $\Gamma$~=~1.77~(1.63$-$1.92), $\chi^2_{\nu}$~=~0.95, $F_{\rm{abs,~0.8-8.0~keV}}$~=~2.2$\times$~10$^{-12}$~erg~cm$^{-2}$~s$^{-1}$, $L_{\rm{0.3-10.0~keV}}$~=~1.4~$\times$~10$^{34}$~$D_{6.1}^2$~erg~s$^{-1}$, and  $L_{\rm{2-10~keV}}$~=~1.1~$\times$~10$^{34}$~$D_{6.1}^2$~erg~s$^{-1}$ (see Figure~\ref{figure:pwnspectrum}, left).

Next, we perform a spatially resolved spectroscopic study of the extended nebula.
To look for variations in the photon index without an artificially varying column density while fitting smaller regions with fewer counts, 
we freeze $N_{\rm{H}}$ to the value from the overall diffuse fit.
Table~\ref{table:spectra} 
shows the results of fitting a power law model to the regions shown in Figure~\ref{figure:pwn} (right) and Figure~\ref{figure:pwnspectrum} (right) shows the power law fit to each region (obs. 11092 and 1037 were fit simultaneously and are grouped together for plotting purposes only).  Beginning at the point source and moving outward across the PWN, we observe an overall increase (or steepening) in the photon index.  This spectral steepening is expected as we move away from a neutron star due to 
expansion and synchrotron losses~\citep{Gaensler2006}.  
We also explored background regions within the diffuse emission for the inner and mid regions.  We find a similar trend in the photon index, however, with a lower net flux and with the parameters being
 less constrained.
The brightening along the edges of the diffuse PWN is well fit by a power law with $\Gamma$~=~1.5~(1.2--1.9) for the eastern arc and $\Gamma$~=~1.7~(1.3--2.1) for the western arc. 

We have searched for thermal emission that would arise from hot plasma in the SNR or interaction with the reverse shock.
We do not find evidence for any significant emission.
In particular, the residuals (lower panels of Figure~\ref{figure:pwnspectrum}) do not show any significant excess at low energies that may indicate the presence of thermal emission in the remnant.  
However, this may be due to the limited sensitivity of the observation and field coverage.
A deeper, more sensitive, X-ray observation covering the entire radio nebula is needed to search for any faint thermal X-ray emission. This is further discussed below.

\section{Discussion}
\label{section:disc}

\subsection{Age}
\label{section:age}
The SNR's age is normally determined using the thermal emission properties of the X-ray emitting plasma associated with the SNR shell. In particular, knowledge of the shock velocity ($v_s$) and the actual SNR size ($R_s$)
yields a direct measurement of the age as: $t$~=~$\eta$~$\frac{R}{v_s}$, where $\eta$ is a factor that depends on the evolutionary phase of expansion of the SNR. 
 For CTB~87, no SNR shell has been detected (as mentioned above).

In the absence of a SNR shell detection, we refer next to the offset between the radio and X-ray emission from the PWN, which can be attributed to the neutron star's motion since its birth following the supernova explosion.
The peak of the 21~cm emission is located at $\alpha$(2000)=20$^{\rm{h}}$16$^{\rm{m}}$04\fs4, $\delta$(2000)=+37\arcdeg12\arcmin31\farcs6, 
$\sim$100\arcsec\ from the X-ray peak (or CXOU J201609.2+371110).  At a distance of $D=6.1D_{6.1}$~kpc~\citep{Kothes2003}, this separation translates to $2.9D_{6.1}$~pc perpendicular to the line of sight.  
Fig. 4b of \citet{Hobbs2005} shows the majority of pulsars with an age $<$3~Myr have a two-dimensional speed in the range $100-500$~km~s$^{-1}$.  From this range, we estimate an age of $\sim$5$-$28~kyr 
for CTB~87, assuming the pulsar was born at the location of the radio peak and is now located at the 
 X-ray peak.   
This is a reasonable assumption particularly since the X-ray emission from a PWN is associated with synchrotron radiation from the freshly injected particles,
 while the radio emission characterizes the older population of synchrotron emitting electrons over the neutron star's lifetime.  Using the average 2D speed of 307 km~s$^{-1}$ for pulsars with an age $<$3~Myr~\citep{Hobbs2005}, we estimate an age of $\sim$9~kyr for CTB~87.
These age estimates are comparable to the 11~kyr age estimated for Vela~X~\citep{LaMassa2008} and the 18~kyr age estimated for G327.1$-$1.1~\citep{Temim2009}, PWNe which we compare CTB~87 to further below.  

The advanced age for CTB~87 (at least an order of magnitude higher than young PWNe like the Crab and G21.5$-$0.9) is also hinted at by the large radio size.
The central 8$^{\prime}$$\times$6$^{\prime}$ radio component implies a physical size of $\sim$(14~$\times$~11)$D_{6.1}$~pc. 
The extended 16$^{\prime}$ diffuse component (Kothes et al. 2003; Kothes et al. in preparation) implies an even larger pulsar wind nebula of $\sim$28$D_{6.1}$~pc in extent. This is slightly smaller, but comparable in
size to 
the unusual PWN G76.9+1.0 (29~pc $\times$ 35~pc) which has been recently argued to be an evolved PWN with an age $<$ 36~kyr and which harbors a newly discovered energetic pulsar with a 9~kyr characteristic
age (Arzoumanian et al. 2011).

In the remainder of this discussion, we will assume an age of $t = 10 t_{10}$~kyr for CTB~87 where calculations require an age estimate, keeping in mind that the age is uncertain given the estimates above (see also an estimate of the putative pulsar's characteristic age in Section~\ref{section:pulsarproperties}).
Moreover, if we take into account the even larger offset (an additional 30$^{\prime\prime}$) between the peak of the \textit{extended} radio emission and the peak of the X-ray emission, 
the age would be 30\% larger for the same assumed velocity. This will be further discussed in the forthcoming radio paper  (Kothes et al. in preparation). 
Additionally, if the offset is due to a reverse shock (Section~\ref{section:reverseshock}) and not related to the proper motion, this estimate may not reflect the age of the system.

We should note that there is accumulating evidence for pulsars' spin axes to be roughly aligned with the direction of their motion, albeit with a small number of systems for which 3D information is available~\citep{Ng2007}.  
\citet{Noutsos2012} provide further evidence that alignment between projected spin-axes and pulsar proper motion is more likely than orthogonality.  
In Figure~\ref{figure:pointsource} (left), we see the possible jets (believed to define the spin-axis) are nearly perpendicular to the apparent proposed proper motion of the putative pulsar. 
If our assumed pulsar's direction of motion and jets' direction are correct, then CTB~87 would be a system that doesn't follow the expected alignment from the above-mentioned studies.
A deeper \textit{Chandra} observation and proper motion studies of the X-ray source will confirm this.

\subsection{Putative Pulsar \& Pulsar Wind Properties}
\label{section:pulsarproperties}
To the best of our knowledge, no pulsations have been reported from CTB~87~\citep{Gorham1996, Biggs1996, Lorimer1998}.  We can however infer the properties of the neutron star powering CTB~87 from the X-ray luminosity of the nebula.  
The unabsorbed flux of CTB~87 in the $2-10$~keV band is $F_{\rm{unabs,~2-10~keV}}=2.4 \times 10^{-12}$~erg~cm$^{-2}$~s$^{-1}$, implying a luminosity of $L_{\rm{X,~2-10~keV}}=1.1\times10^{34}~D_{6.1}^2$~erg~s$^{-1}$.  We can estimate the spin-down energy loss of the neutron star powering the PWN from the PWN luminosity using the empirical relationship $L_{\rm{X,~2-10~keV}}=10^{-19.6} \dot{E}^{1.45}$~\citep{Li2008}.  This yields an estimate of $\dot{E}=9.6 \times 10^{36}~D_{6.1}^{1.38}$~erg~s$^{-1}$, consistent with the estimate of $\dot{E}=6.9 \times 10^{36}~D_{6.1}^{1.49}$~erg~s$^{-1}$ obtained with the empirical relationship $L_{\rm{X,~2-10~keV}}=1.34 \log \dot{E} -15.34$~\citep{Possenti2002}.  The efficiency in converting the spin-down luminosity to synchrotron emission is then $\eta_X \equiv L_X/\dot{E} \approx 0.001$, which is within the range observed for PWNe powered by rotation-powered pulsars.  
The PWN luminosity can also be used to estimate the characteristic age ($\tau$) of the neutron star using $L_{\rm{X,~2-10~keV}}=10^{42.4} \tau^{-2.1}$~\citep{Li2008}.  We estimate an age of $\tau=9800~D_{6.1}^{-0.95}$~years.

Assuming an age of $t~=~10t_{10}$~kyr (keeping in mind the large uncertainty on this estimate, noted in Section~\ref{section:age}), $\dot{E} \sim 10^{37} \dot{E}_{37}$~erg~s$^{-1}$, a braking index of $n$=3 (magneto-dipole braking), a moment of inertia of $I$~=~10$^{45} I_{45}$~g~cm$^2$, and a neutron star radius of $R$~=~10$R_{10}$~km, we can estimate the properties of the putative pulsar~\citep[equations described for e.g. in][]{Gaensler2006}.  The period of the pulsar is estimated as $P=\left( \frac{4 \pi^2 I}{(n-1)\dot{E}\tau} \right)^{1/2}$ $\approx 0.08~I_{45}^{1/2}~\dot{E}_{37}^{-1/2}~t_{10}^{-1/2}$~seconds.  The corresponding period derivative is $\dot{P} = \frac{P}{(n-1)\tau} \approx 1.3 \times 10^{-13}~I_{45}^{1/2}~\dot{E}_{37}^{-1/2}~t_{10}^{-3/2}$ and the corresponding equatorial surface dipole magnetic field is $B = 3.2~\times~10^{19}~I_{45}^{1/2}~R_{10}^{-3}~( P \dot{P} )^{(1/2)} \approx 3.2 \times 10^{12}~I_{45}~\dot{E}_{37}^{-1/2}~R_{10}^{-3}~t_{10}^{-1}$~G, 
within the range of the field expected for pulsars with prominent PWNe~\citep[$1\times10^{12} - 5\times10^{13}$~G,][]{Gaensler2006}.

We expect to observe a termination shock around a rotation-powered neutron star, at a radius $r_t=\sqrt{\dot{E}/(4 \pi c P_{\rm{PWN}})}$; where $P_{\rm{PWN}}$ is the nebular pressure.
 Figure~\ref{figure:pointsource} suggests that the arc seen at a radius of $\sim$5\arcsec\  corresponds to the termination shock.  
 At an assumed distance of 6.1~kpc, this corresponds to a distance of 0.15~pc from the point source.    
 This estimate is in agreement with the typical scale of 0.1~pc observed in other PWNe~\citep{Gaensler2006}.  
Using the spin-down energy inferred above, and equating the expression of the termination shock site to 5\arcsec,  we estimate the pressure in the PWN to be $P_{\rm{PWN}}\sim 1.2 \times 10^{-10} D_{6.1}^{-0.62}$~dyne~cm$^{-2}$.  We can then estimate the magnetic field in the nebula as $B_n=(8 \pi P_{\rm{PWN}})^{1/2}\sim 55 D_{6.1}^{-0.31}~\mu$G.
This estimate is in agreement with the preliminary value inferred from the radio study assuming equipartition between particle energy and magnetic field energy (Kothes et al. in preparation).

\subsection{Ambient Density \& the Missing SNR Shell}
The absence of a SNR shell detection implies expansion into a low-density medium.
To infer the upper limit on the 
density of the medium that the pulsar wind nebula is expanding into (which would be the shocked ejecta for a Vela-like SNR),
we estimate next the maximum thermal contribution to the diffuse emission allowed by the data by adding a thermal component to the power law model. 
 We freeze the best fit power law parameters (Table~\ref{table:spectra}, `Total Diffuse' region), and add an Astrophysical Plasma Emission Code (APEC) thermal component~\citep{Smith2001}.   
We increase the emission measure ($EM$) until the model level under 1~keV exceeds the data by 2$\sigma$, finding a conservative upper limit for the normalization on the thermal component of 
9~$\times$~10$^{-4}$~cm$^{-5}$ (for $kT$~=~0.5~keV).  
This implies that $\int n_e n_{\mathrm H} dV \sim f n_e n_{\mathrm H} V$~=~10$^{14}$ $\left( 4 \pi D^2\right) \left( 9~\times~10^{-4}~\rm{cm}^{-5} \right)$~$<$~4.0~$\times$~10$^{56}$~$D^2_{6.1}$~cm$^{-3}$, 
where $f$ is the volume filling factor, $n_e$ is the post-shock
electron density, and $n_{\mathrm H} \sim n_e / 1.2$.  Using a volume
of 5.6~$\times$~10$^{57}$~$D^3_{6.1}$~cm$^3$ for the PWN, $n_e < 0.29 f^{-1/2}~D^{-1/2}_{6.1}$~cm$^{-3}$.   
If a SNR shell exists beyond the detected PWN, the shocked ISM must have a density lower than that estimated above for the ejecta, $0.29 f^{-1/2}~D^{-1/2}_{6.1}$~cm$^{-3}$ = $0.92~D^{-1/2}_{6.1}$~cm$^{-3}$ (for a volume filling factor $f \sim 0.1$).  This then implies an upper limit on the upstream ambient density of $n_0 < 0.2~D^{-1/2}_{6.1}$~cm$^{-3}$ ($\rho_0 < 3 \times 10^{-25}~D^{-1/2}_{6.1}$~g~cm$^{-3}$ where $n_0$ includes only hydrogen), assuming a compression ratio of 4.  The inferred 
low density supports the lack of detection of the SNR shell or X-ray emitting ejecta,
and is suggestive of expansion into a stellar wind-blown bubble \citep{Weaver1977}.   
This upper limit is consistent with estimates obtained by extracting the background spectrum (outside the PWN) and assuming that the thermal emission from a putative shell is buried underneath the background.  From these background spectra, we estimate an upper limit of $9 \times 10^{-14}$~erg~cm$^{-2}$~s$^{-1}$~arcmin$^{-2}$ on the observed X-ray flux of the putative SNR shell.

The above density upper limit is similar to density estimates for G21.5-0.9~\citep{SafiHarb2001, Matheson2010}, whose shell was revealed after accumulating $\sim$0.5~Msec using \textit{Chandra}.  A deep X-ray observation of CTB~87 to search for the missing SNR shell is therefore warranted.
Additionally, future studies of the surrounding environment with CO or HI data may help to further constrain the density.  This scenario will be revisited and further discussed in the follow-up radio paper (Kothes et al. in preparation).

Assuming that the remnant is evolving in a uniform medium, the reverse shock will reach the center of the SNR in a time $t_{\rm{Sedov}} \approx 7 (M_{ej}/10~M_{\sun})^{5/6} (E_{\rm{SN}}/10^{51}~\rm{ergs})^{-1/2} (n_0/1~\rm{cm}^{-3})^{-1/3}$~kyr.  For an ejected mass of 
a few solar masses, an explosion energy of $10^{51}$~ergs, and a number density of $n_0 < 0.2$~cm$^{-3}$, this is greater than 4~kyr for CTB~87, consistent with our estimate above for the age (see Section 5.1).

For a remnant in the Sedov stage, the SNR radius can be estimated as $R_{\rm{SNR}} \approx 1.15 \left( \frac{E_{\rm{SNR}}}{\rho_{\rm{ISM}}} \right)^{1/5} t^{2/5}$~\citep{vanderSwaluw2003}.  
Again assuming an explosion energy of $E_{\rm{SNR}} = 10^{51} E_{51}$~erg, an ISM density of $n_0 = 0.2~n_{0.2}$~cm$^{-3}$, and an age of $t = 10 t_{10}$~kyr (large uncertainty noted in Section~\ref{section:age}), we estimate the SNR radius to be $R_{\rm{SNR}} \approx 18 \left( \frac{E_{51}}{n_{0.2}} \right)^{1/5} t_{10}^{2/5}$~pc, 
which is larger than the revised size of the PWN ($\sim$28~pc diameter at 6.1~kpc; see Section 5.1).
We conclude that all of the above mentioned arguments point to CTB~87 being an evolved PWN expanding into a low density medium, likely due to the stellar wind bubble blown by the progenitor star.

\subsection{Morphology Interpretation}
In Section 5.2 we have interpreted the compact $\sim$5\arcsec\ nebula shown in Figure~2 as a wind-blown bubble associated with the deposition of the neutron star's wind energy into the surroundings.
As shown in Figure~1, in addition to this compact nebula, the X-ray emission from CTB~87 is characterized by a cometary-like structure extending $\sim$250\arcsec\  northwest from CXOU~J201609.2+371110
and bridging the X-ray and radio emission peaks.
As mentioned earlier, this extended emission has some brightening along its edges (see Figure~1). 
This morphology may be interpreted in two ways.  First, these arcs may be a bow shock structure due to the motion of the pulsar to the southeast through the SNR interior, leading to the formation of
a PWN with a hybrid morphology (i.e. a wind-blown bubble and a bow shock nebula).  
Alternatively, the reverse SNR shock may be running into, and crushing, the PWN.  We discuss next these two scenarios.

\subsubsection{Bow Shock Interpretation}

A pulsar given a kick at birth that has sufficient velocity will move out of the PWN and create a bow shock.  As the pulsar moves outward in the SNR, away from the supernova explosion site, the sound speed in the shocked ejecta decreases and the pulsar's motion becomes supersonic.  The ram-pressure from the pulsar's motion confines the nebula and the PWN evolves to a cometary appearance, with the bow shock forming around the head of the PWN after the pulsar has crossed $\sim$68\% of the SNR shell, at half of the crossing time~\citep{vanderSwaluw2004,Gaensler2006}.

The pulsar wind in a bow shock PWN decelerates at a termination shock where the pressure balance is due to ram pressure from the neutron star's motion.  In the direction of the star's motion, the termination shock radius, $R_{\omega 0}$, is defined by 
$\dot{E}/\left(4 \pi \omega R^2_{\omega 0} c\right) = \rho_0 v^2_{\mathrm{PSR}}$ ($\omega = 1$ for an isotropic wind), and is located at approximately half of the observed bow shock radius~\citep{Gaensler2006}.  

Ram-pressure confinement in this hybrid morphology scenario requires that the ram pressure due to the pulsar's proper motion in the SNR environment,  $\rho_0 v^2_{\mathrm{PSR}}$, exceeds the nebular pressure, $P_{\mathrm{PWN}}$, estimated in Section~5.2 ($\rho_0$ is the density of the medium in which the PSR is moving). 
Assuming that the brightening along the PWN edge in Figure~\ref{figure:pwn} is due to a bow shock, $R_{\omega 0} \sim 1.8 \times 10^{18}~D_{6.1}$~cm.
 Using our inferred value for $\dot{E}$ (Section 5.2), we find that $\dot{E}/\left(4 \pi \omega R^2_{\omega 0} c\right)~\sim~8 \times 10^{-12}$~dyne~cm$^{-2}$. 
This is only $\sim$7\% of the pressure $P_{\mathrm{PWN}} = 1.2 \times 10^{-10}$~dyne~cm$^{-2}$ estimated from the termination shock of the more compact nebula (Section~5.2), a ratio that is consistent with $(\frac{r_t}{R_{\omega 0}})^2$ .
 Furthermore, for  $\rho_0 v^2_{\mathrm{PSR}}$  to exceed this nebular pressure of $\sim$10$^{-10}$~dyne~cm$^{-2}$, $v_{\mathrm{PSR}}$ should exceed $\sim$170~km~s$^{-1}$ (for an assumed ambient density of 0.2~cm$^{-3}$, Section 5.3).    
As the formation of a prominent bow shock requires the nebular pressure to be exceeded by a large amount, the above suggests a high pulsar velocity is needed for creation of a bow shock and argues against the cometary morphology being caused by ram-pressure confinement.

\subsubsection{Reverse Shock Interaction Interpretation}
\label{section:reverseshock}
In the following, we discuss the observed morphology in the light of an evolved PWN encountering the SNR reverse shock. This is motivated by the age inferred above (Section~5.1) and the offset observed between the radio
and X-ray nebulae, as seen in other evolved PWNe further described below.  
A detailed description of the evolution of a PWN in a SNR can be found in \citet{Blondin2001},  \citet{vanderSwaluw2004}, and \citet{Gaensler2006}. 
We here briefly summarize.

As the expanding SNR sweeps up sufficient mass from the ISM, it evolves into a Sedov-Taylor phase. A reverse shock develops, running back into the supernova ejecta and eventually crushing the PWN, causing it to rebound and oscillate several times.
The timescale for the reverse shock to interact with the entire PWN is a significant fraction of the PWN life~\citep{vanderSwaluw2004}.  After these oscillations fade, the PWN again expands, now into hot, shocked ejecta.  The pulsar may have traveled a distance comparable or larger than the equivalent spherical PWN around a stationary pulsar.  The pulsar leaves its original wind bubble and generates a new smaller PWN at its current location, off-center within the PWN.  
Therefore, after the passage of the reverse shock the PWN consists of (i) a \textit{relic} PWN from the initial energetic stage of the pulsar wind, appearing as a central (possibly distorted) radio PWN with little corresponding X-ray emission, and (ii) a head containing the pulsar, directed toward the SNR shell, and emitting in X-rays.

\Citet{vanderSwaluw2004} express an upper limit on the age, $t_{col}$, at which the reverse shock collides with the entire PWN surface as $t_{col} = 1045 E_{51}^{-1/2} \left( \frac{M_{ej}}{M_{\sun}} \right)^{5/6} n_0^{-1/3}$~yr where $E_{51}$ is the total mechanical energy of the SNR in units of $10^{51}$~erg, $M_{ej}$ is the ejected mass, and $n_0$ is the ambient hydrogen number density.  For CTB~87 we assume (as above) the explosion energy is $10^{51}$~erg, that the ejected mass is 
a few solar masses, and that the ambient density is $<$0.2~cm$^{-3}$. We find that $t_{col} > 4$~kyr for CTB~87, at the low end of the estimated age (Section~5.1).
This suggests that the reverse shock has potentially encountered the PWN, and is the cause of the observed morphology. 
 The supernova explosion may also have been asymmetric, or the ambient density may be non-uniform, leading to a more rapid evolution in one direction.
 This is to be further discussed in the forthcoming radio paper (Kothes et al., in preparation).

This conclusion is further supported by the TeV emission detected from CTB~87~\citep{Aliu2011}.  
Focusing on \textit{Chandra}-detected PWNe near extended TeV sources,  \citet{Kargaltsev2008} find that faint extensions of compact PWNe point toward offset centers of TeV emission, supporting the notion that these TeV sources are crushed PWNe.  Recent \textit{VERITAS} studies of CTB~87 indicate that the TeV emission overlaps with the radio emission, further supporting the picture of a `relic' PWN (E. Aliu and M. Roberts 2012, private communication).
The lack of X-ray emission from the SNR ejecta or SNR shell remains however a question to be addressed with deeper observations.

\subsubsection{Comparison to other evolved PWNe}
G327.1$-$1.1 is a composite remnant with a symmetric radio shell and off-center non-thermal component (PWN), similar in structure to CTB~87.  There is a radio finger extending from the PWN, at the end of which is a compact X-ray source that may be the pulsar~\citep{Temim2009}.  \chan\ images show a compact source embedded in a cometary structure that trails back toward the radio PWN, as well as two prong-like structures extending into a large bubble in the opposite direction from the PWN~\citep{Temim2009}.  
We see a similar structure in CTB~87 (Figure 1) with the X-ray nebula offset from the radio nebula and most of the X-ray nebula trailing back toward the radio nebula.  We also see emission ahead of the point source in CTB~87, extending to the southeast away from the radio nebula.
\citet{Temim2009} similarly favor a reverse shock interaction scenario over a bow shock scenario.
G327.1$-$1.1 however has more detailed spectra available, and thermal X-ray emission has been detected with \textit{XMM-Newton}.
The SNR shell detection in G327.1$-$1.1 also allows further constraints to be placed on the SNR properties.  
TeV emission from G327.1$-$1.1 has been detected by \textit{H.E.S.S.}, showing the $\gamma$-ray centroid is located between the radio centroid and the pulsar candidate~\citep{Acero2011}.

Vela-X (associated with PSR~B0833-45) is another example of a PWN whose morphology has been affected by its interaction with the reverse shock (Blondin et al. 2001).  The radio nebula extends south of the pulsar, the northern portion possibly having been disturbed by the reverse shock.  Vela-X has a complicated filamentary structure in X-rays, with thermal emission and enhanced O, Ne, and Mg abundances detected, signifying ejecta that has been mixed into the nebula by the propagation of the reverse shock~\citep{LaMassa2008}. 
The explanation of Vela-X as a relic PWN was confirmed by \citet{Aharonian2006} with the detection of TeV emission that is offset from the pulsar.

\section{Conclusions}

We have presented the first detailed X-ray study of CTB~87, a large radio nebula classified as a PWN.  Thanks to \textit{Chandra}'s unprecedented resolution, we have detected a point source, CXOU~J201609.2+371110, the likely putative pulsar powering CTB~87,
inside a compact X-ray nebula that is $\sim$100\arcsec\ offset from the peak of the radio nebula. 
The compact $\sim$5\arcsec-radius nebula is characterized by a torus/jet-like structure indicating its origin as the putative pulsar's wind-blown bubble.
In addition to the X-ray source and compact nebula, a more extended nebula was found to
trail back toward the peak of the radio emission.
A spatially resolved spectroscopic study of the nebula shows that the PWN is well fit by a power law model, with the overall spectrum steepening away from the point source.  We also observed brightening at the edges of the PWN, and discussed two possible interpretations:
a bow shock caused by the motion of the pulsar or
a PWN crushed by the SNR's reverse shock. 
Based on the observed multi-wavelength morphology, timescale for the reverse shock to interact with the PWN, and estimated age of the PWN,
the reverse shock interaction scenario seems more plausible. 
A deep X-ray observation is needed to search for thermal emission associated with the shocked ejecta and/or the SNR shell, as well as to confirm the high-resolution structures detected in the compact nebula.
Timing studies in the radio and X-ray will allow the search for pulsations from
CXOU~J201609.2+371110. Finally, in the long term, a proper motion study of the putative pulsar is needed to shed light on its motion within the radio nebula.

\acknowledgments

We thank the referee for useful comments on the manuscript. 
The research presented in this paper has used data from the Canadian Galactic Plane Survey, a Canadian project with international partners, supported by the Natural Sciences and Engineering Research Council (NSERC),
and made use of the NASA's Astrophysics Data System and the High-Energy Astrophysics Science Archive Research maintained at NASA's Goddard Space Flight Center.
This research was primarily supported by NSERC. SSH acknowledges support from an NSERC Discovery Grant,  the Canada Research Chairs program, the Canadian Space Agency, the Canadian Institute for
Theoretical Astrophysics, and Canada's Foundation for Innovation.

{\it Facilities:} \facility{ASCA (GIS)}, \facility{CXO (ACIS)}, \facility{DRAO:Synthesis Telescope}.

\clearpage

\clearpage

\clearpage

\begin{deluxetable}{llrcr}
\tabletypesize{\scriptsize}
\tablecaption{X-ray Observations of CTB~87}
\tablewidth{0pt}
\tablehead{
\colhead{Satellite} & \colhead{Obs. Date} & \colhead{ObsID} & \colhead{PI} & \colhead{Effective Exposure (ks)}
}
\startdata
$ASCA$ (GIS) & 1995 May 29 & 53041000 & Gehrels & 53.1\\
$Chandra$ (ACIS-S) & 2001 July 8 & 1037 & Garmire & 17.8\\
$Chandra$ (ACIS-I) & 2010 January 16 & 11092 & Safi-Harb & 69.3
\enddata
\label{table:observations}
\end{deluxetable}

\begin{deluxetable}{lcccccc}
\tabletypesize{\scriptsize}
\rotate
\tablecaption{Power Law Fit Results for \chan\ Data}
\tablewidth{0pt}
\tablehead{
\colhead{Parameter} & \colhead{0\arcsec$-$2\arcsec} & \colhead{2\arcsec$-$10\arcsec} & \colhead{Inner Diffuse} & \colhead{Mid Diffuse} & \colhead{Outer Diffuse} & \colhead{Total Diffuse}
}
\startdata
\# Observed Counts (0.3$-$10.0 keV)      & 321                    & 822                 & 4390               & 4400             & 13 677             & 22473\\  \# Counts After Bkgd Subtraction        & 202                    & 578                 & 3195               & 2439            & 3767                & 9356\\  
Minimum \# Counts per bin                          & 5                      & 10                  & 50                & 50               & 50                  & 50\\
Region Area (arcsec$^2$)                & 12.6                   & 301.6               & 6.31E3             & 1.03E4           & 5.23E4              & 6.92E4\\
\tableline
$N_{\rm{H}}$ (10$^{22}$~cm$^{-2}$)               &  1.4 (frozen)          & 1.4 (frozen)        & 1.4 (frozen)       & 1.4 (frozen)     & 1.4 (frozen)        & 1.38 (1.21 $-$ 1.57)\\
Photon Index, $\Gamma$                  &  1.1 (0.7 $-$ 1.6)       & 1.2 (0.9 $-$ 1.4)     & 1.57 (1.50 $-$ 1.65)    & 1.72 (1.63 $-$ 1.82)  & 1.83 (1.69 $-$ 1.96)     & 1.68 (1.54 $-$ 1.84)\\
Norm. (photons~keV$^{-1}$~cm$^{-2}$~s$^{-1}$) & 5.2 (2.8$-$8.3) E-6  & 1.6 (1.2$-$2.0) E-5   & 1.58 (1.46$-$1.71) E-4  & 1.39 (1.25$-$1.53) E-4 & 2.3 (2.0$-$2.6) E-4  & 5.1 (4.2 $-$ 6.3) E-4\\
$\chi^2_{\nu}$                           & 0.91                   & 1.10                 & 0.92              & 1.22              & 1.04               & 1.01\\
$\nu$                                   & 55                     & 70                  & 78                & 67                 & 161                & 314\\
Abs. Flux (erg~cm$^{-2}$~s$^{-1}$, 0.3$-$10.0 keV)        & 5.2E-14                & 1.5E-13             & 7.7E-13           & 5.4E-13            & 7.4E-13            & 2.1E-12\\
Luminosity ($D_{6.1}^2$~erg~s$^{-1}$, 0.3$-$10.0 keV)                & 3.0E32                 & 8.7E32              & 5.5E33            & 4.2E33             & 6.3E33             & 1.6E34\\
Luminosity ($D_{6.1}^2$~erg~s$^{-1}$, 2.0$-$10.0 keV)                & 2.5E32                 & 6.5E32              & 3.5E33            & 2.5E33             & 3.4E33             & 9.6E33 
\enddata
\tablecomments{$N_{\rm{H}}$ was frozen to 1.4$\times$10$^{22}$~cm$^{-2}$, the best fit to the entire diffuse emission.  All confidence ranges are 90\%.  Luminosity assumes a distance of 6.1~kpc.  The total X-ray luminosity of CTB~87 is 1.7$\times$10$^{34}$~erg~s$^{-1}$.  See Figure~\ref{figure:pwn} (right) for the inner, mid, and outer regions.}
\label{table:spectra}
\end{deluxetable}

\end{document}